УДК 621.3.049.77::621.3.082.61::536.413.2

**Л. С. Синев, В. Т. Рябов**

**Расчет коэффициентных напряжений в соединениях кремния со стеклом**

Рассмотрены напряжения, возникающие в деталях из стекла и кремния после их соединения электростатическим методом из-за разницы коэффициентов теплового линейного расширения материалов. Представлено описание способа расчета напряжений в сборке в соответствии с теорией слоистых композитов. Приведены расчетные распределения коэффициентных напряжений внутри соединенных материалов разных толщин.

Ключевые слова: анодная посадка, электростатическое соединение, тепловое расширение, напряжение.

**Sinev L. S., Ryabov V. T.**
**COEFFICIENT OF THERMAL EXPANSION MISMATCH INDUCED STRESS CALCULATION FOR FIELD ASSISTED BONDING OF SILICON TO GLASS**

The residual stress induced in assembly is a common concern in electronic packaging. The mismatch in coefficient of thermal expansion between borosilicate glass and silicon, upon temperature variation, generates an internal stress state. This affects important characteristics of microelectromechanical devices or constituent elements. Such as self frequence or stiffness. Stresses caused by thermal expansion coefficients mismatch of anodically bonded glass and silicon samples are studied in this paper. Stress calculation based on lamination theory is presented. Usage examples of such calculations are described. For bonded silicon and LK-5 glass several results of calculations are presented. Stress distribution in bonded silicon and glass of several thicknesses is evaluated. Stress distribution in bonded glass-silicon-glass structure is evaluated. Bonded silicon surface stress dependence of glass to silicon wafer thickness ratio is evaluated. Experimental study of thermal mismatch stress in glass based on birefringence phenomenon was conducted. It's results are presented in this paper.

**Введение**

Микросистемная техника (МСТ) является в настоящее время одним из наиболее развивающихся междисциплинарных научно-технических направлений. Стремительное развитие технологий микросистемной техники за последние годы показало огромный потенциал этой области и позволило реализовать множество элементов, которые невозможно изготовить с использованием макротехнологий.

На выходные характеристики приборов микросистемной техники заметно влияют напряжения, привнесенные использованными материалами. Такие напряжения могут вызывать изменение формы всего прибора или его отдельных элементов, а также влиять на рабочие характеристики, такие как жесткость или собственная частота. Одними из наиболее распространенных материалов микросистемной техники являются кремний и стекло.

Электростатическое соединение является одной из основных сборочных операций микросистемной техники. В данном процессе кремний соединяется со стеклом посредством приложения внешней разности потенциалов и одновременного нагрева до температур 200…450 °C [1, 2].

В результате соединения образуются коэффициентные напряжения, то есть напряжения, возникающие вследствие разности значений коэффициентов теплового линейного расширения (КТЛР) стекла и кремния [3].

До начала нагрева стеклянная и кремниевая детали имеют одинаковые размеры, при нагреве детали расширяются неравномерно и при температуре соединения $T_с$ имеют отличающиеся размеры. После соединения детали, охлаждаясь до рабочей температуры $T_\text{р}$, взаимно деформируются.

Целью данной работы является определение соотношения КТЛР стекла и кремния для обеспечения соединения с минимальными коэффициентными напряжениями. Нелинейная зависимость КТЛР соединяемых деталей от температуры не позволяет минимизировать коэффициентные напряжения путем подбора материалов с близкими средними КТЛР.

**Методика расчета**

Истинным коэффициентом теплового линейного расширения называется отношение изменения линейного размера тела, деленного на его начальный размер, к малому изменению температуры, вызвавшему изменение размера тела [4]:

$$\alpha = \frac{1}{l_0} \cdot \frac{dl}{dT},$$

где $\alpha$ — истинный коэффициент теплового расширения, 1/°С; $l_0$ — начальный размер тела, м; $dl$ — изменение линейного размера тела, м; $dT$ — малое изменение температуры, вызвавшее изменение размера тела, °С.

Воспользуемся теорией слоистых композитов [5]. Рассмотрим соединенные детали как многослойный композиционный материал. В качестве координатной плоскости $xy$ примем срединную плоскость пластины, то есть плоскость, лежащую до нагружения пластины посередине между ее верхней и нижней поверхностями. Считаем, что эта плоскость не меняет своего положения в процессе нагружения. Положительным направлением оси $z$ будем считать направление вниз. Пластина состоит из произвольного числа слоев, соединенных друг с другом. Для каждого слоя справедлив закон Гука. Предполагаем, что слои не оказывают сдавливающего воздействия один на другой. Толщину пластины считаем неизменной. Используем допущения, что нагрев пластины равномерен, что длина и ширина пластины значительно превышают её толщину. Изменение жесткости рассматриваемых материалов считаем незначительным. Влияние краевых эффектов и разницы в коэффициентах теплопроводности материалов исключаем из рассмотрения. Также не учитываем изменение свойств стекла, связанное с переносом ионов в результате проведения процесса электростатического соединения [6]. Положительными напряжениями считаем напряжения растяжения в материале.

На рисунке 1 представлена иллюстрация применяемой модели слоистого композита. В этой модели $t$ — толщина многослойной пластины; 1, 2, …$k$, … $n$ — номер слоя.

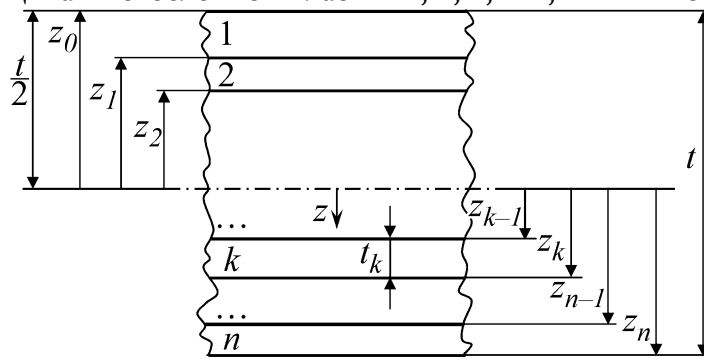

Рис. 1. Иллюстрация модели слоистого композита.

Упругие свойства каждого слоя могут быть представлены при помощи матриц коэффициентов жесткости.

$$\mathbf{C_{Si}} = \begin{pmatrix} 167{,}4 & 65{,}2 & 65{,}2 & 0 & 0 & 0 \\ 65{,}2 & 167{,}4 & 65{,}2 & 0 & 0 & 0 \\ 65{,}2 & 65{,}2 & 167{,}4 & 0 & 0 & 0 \\ 0 & 0 & 0 & 79{,}6 & 0 & 0 \\ 0 & 0 & 0 & 0 & 79{,}6 & 0 \\ 0 & 0 & 0 & 0 & 0 & 79{,}6 \end{pmatrix},$$

$$\mathbf{C_g} = \begin{pmatrix} 74{,}6 & 16{,}8 & 16{,}8 & 0 & 0 & 0 \\ 16{,}8 & 74{,}6 & 16{,}8 & 0 & 0 & 0 \\ 16{,}8 & 16{,}8 & 74{,}6 & 0 & 0 & 0 \\ 0 & 0 & 0 & 28{,}9 & 0 & 0 \\ 0 & 0 & 0 & 0 & 28{,}9 & 0 \\ 0 & 0 & 0 & 0 & 0 & 28{,}9 \end{pmatrix},$$

где $\mathbf{C_{Si}}$, $\mathbf{C_g}$ — матрицы коэффициентов жесткости кремния ориентации (100) (в ГПа) [7] и стекла ЛК-5 (в ГПа) [8, 9].

Уравнение для напряжений в каждом слое при механическом и тепловом нагружении:

$$\boldsymbol{\sigma} = \mathbf{Q}\left(\boldsymbol{\varepsilon} - \boldsymbol{\varepsilon}^T\right), \qquad (1)$$

$$\boldsymbol{\varepsilon}^T = \int_{T_\text{с}}^{T_\text{р}} \boldsymbol{\alpha}(T)\,\mathrm{d}T,$$

$$\boldsymbol{\varepsilon} = \boldsymbol{\varepsilon}^0 + z\boldsymbol{\kappa},$$

где $\boldsymbol{\sigma}$ — вектор напряжений, Па; $\mathbf{Q}$ — преобразованная матрица жесткости каждого слоя, Па; $\boldsymbol{\varepsilon}$ — вектор индуцированных деформаций (растяжения), вызванных механической нагрузкой; $\boldsymbol{\varepsilon}^T$ — вектор индуцированных деформаций (растяжения), вызванных тепловой нагрузкой; $T_\text{с}$ — температура соединения, °C; $T_\text{р}$ — рабочая температура, °C; $\boldsymbol{\varepsilon}^0$ — относительное удлинение срединной поверхности многослойной композитной пластины (по осям); $\boldsymbol{\kappa}$ — радиус кривизны срединной поверхности многослойной композитной пластины (по осям), 1/м; $z$ — расстояние, измеряемое от срединной поверхности, м. Матрицы $\mathbf{Q}$ получены из матриц жесткости в соответствии с формулами поворота системы координат [10] так, чтобы они отражали упругие свойства слоев в принятых нами направлениях координатных осей.

Поскольку мы приняли допущение, что размеры пластины много больше её толщины, то далее будем рассматривать напряженное состояние многослойной пластины как плоское напряженное состояние [10]. Таким образом, векторы и матрицы в уравнениях будут представлены в сокращенной форме за счет отбрасывания незадействованных компонентов (сокращения до размера 3 × 3).

Учитывая вышесказанное, уравнение (1), может быть записано в матричном представлении следующим образом:

$$\begin{pmatrix} \sigma_x \\ \sigma_y \\ \tau_{xy} \end{pmatrix} = \begin{pmatrix} Q_{11} & Q_{12} & Q_{16} \\ Q_{12} & Q_{22} & Q_{26} \\ Q_{16} & Q_{26} & Q_{66} \end{pmatrix} \left( \begin{pmatrix} \varepsilon_x^0 \\ \varepsilon_y^0 \\ \gamma_{xy}^0 \end{pmatrix} + z \begin{pmatrix} \kappa_x \\ \kappa_y \\ \kappa_{xy} \end{pmatrix} - \begin{pmatrix} \varepsilon_x^T \\ \varepsilon_y^T \\ \gamma_{xy}^T \end{pmatrix} \right),$$

где $\sigma_x$, $\sigma_y$, $\tau_{xy}$ — элементы вектора напряжений, Па; $Q_{ij}$ — элементы преобразованной матрицы жесткости каждого слоя, Па; $\varepsilon_x^0$, $\varepsilon_y^0$, $\gamma_{xy}^0$ — элементы вектора относительного удлинения срединной поверхности многослойной композитной пластины; $\kappa_x$, $\kappa_y$, $\kappa_{xy}$ — элементы векторного представления радиуса кривизны срединной поверхности многослойной композитной пластины, 1/м; $\varepsilon_x^T$, $\varepsilon_y^T$, $\gamma_{xy}^T$ — элементы вектора индуцированных деформаций, вызванных тепловой нагрузкой.

Результирующие силы и моменты (обобщенные силовые факторы), воздействующие на пластину, определяются посредством интегрирования уравнения (1) по всей толщине пластины:

$$\mathbf{N} = \int_t \boldsymbol{\sigma} \, dz,$$

$$\mathbf{M} = \int_t \boldsymbol{\sigma} \, z \, dz,$$

где $t$ — толщина многослойной пластины, м; $\mathbf{N}$ — результирующая нагрузка, отнесенная к единице длины линий, ограничивающих элемент рассматриваемой поверхности [11], Н/м; $\mathbf{M}$ — результирующий момент, отнесенный к единице длины линий, ограничивающих элемент рассматриваемой поверхности, Н.

В рассматриваемом случае, когда жесткость каждого слоя $\mathbf{Q}$ неизменна по толщине слоя, можно записать:

$$A_{ij} = \sum_{k=1}^{n} (Q_{ij})_k (z_k - z_{k-1}),$$

$$B_{ij} = \frac{1}{2} \sum_{k=1}^{n} (Q_{ij})_k (z_k^2 - z_{k-1}^2),$$

$$D_{ij} = \frac{1}{3} \sum_{k=1}^{n} (Q_{ij})_k (z_k^3 - z_{k-1}^3),$$

где $A_{ij}$ — элементы матрицы жесткости при растяжении (мембранной жесткости) [5, 11], Н/м; $B_{ij}$ — элементы матрицы жесткости изгиб-растяжение (смешанной жесткости), Н; $D_{ij}$ — элементы матрицы жесткости при изгибе (изгибной жесткости), Н·м; $z_k$ — расстояние до текущего слоя, измеряемое от срединной поверхности (см. рис. 1).

Тогда взаимосвязь нагрузок и деформаций может быть представлена в уравнениях в матричной форме:

$$\begin{pmatrix} N_x \\ N_y \\ N_{xy} \end{pmatrix} = \begin{pmatrix} A_{11} & A_{12} & A_{16} \\ A_{12} & A_{22} & A_{26} \\ A_{16} & A_{26} & A_{66} \end{pmatrix} \begin{pmatrix} \varepsilon_x^0 \\ \varepsilon_y^0 \\ \gamma_{xy}^0 \end{pmatrix} + \begin{pmatrix} B_{11} & B_{12} & B_{16} \\ B_{12} & B_{22} & B_{26} \\ B_{16} & B_{26} & B_{66} \end{pmatrix} \begin{pmatrix} \kappa_x \\ \kappa_y \\ \kappa_{xy} \end{pmatrix} - \begin{pmatrix} N_x^T \\ N_y^T \\ N_{xy}^T \end{pmatrix}, \quad (2)$$

$$\begin{pmatrix} M_x \\ M_y \\ M_{xy} \end{pmatrix} = \begin{pmatrix} B_{11} & B_{12} & B_{16} \\ B_{12} & B_{22} & B_{26} \\ B_{16} & B_{26} & B_{66} \end{pmatrix} \begin{pmatrix} \varepsilon_x^0 \\ \varepsilon_y^0 \\ \gamma_{xy}^0 \end{pmatrix} + \begin{pmatrix} D_{11} & D_{12} & D_{16} \\ D_{12} & D_{22} & D_{26} \\ D_{16} & D_{26} & D_{66} \end{pmatrix} \begin{pmatrix} \kappa_x \\ \kappa_y \\ \kappa_{xy} \end{pmatrix} - \begin{pmatrix} M_x^T \\ M_y^T \\ M_{xy}^T \end{pmatrix}, \quad (3)$$

где $N_{ij}$ — элементы вектора результирующей нагрузки, отнесенной к единице длины линий, ограничивающих элемент рассматриваемой поверхности, Н/м; $M_{ij}$ — элементы вектора результирующего момента, отнесенного к единице длины линий, ограничивающих элемент рассматриваемой поверхности, Н; $N_{ij}^T$ — элементы векторной записи усилия, вызванного тепловым воздействием, отнесенного к единице длины линий, ограничивающих элемент рассматриваемой поверхности, Н/м; $M_{ij}^T$ — элементы векторной записи момента силы, вызванного тепловым воздействием, отнесенного к единице длины линий, ограничивающих элемент рассматриваемой поверхности, Н.

Силы и моменты, вызванные тепловым воздействием, определяются следующими уравнениями.

$$\mathbf{N}^T = \int_t \mathbf{Q}\varepsilon^T \, dz, \tag{4}$$

$$\mathbf{M}^T = \int_t \mathbf{Q}\varepsilon^T z \, dz, \tag{5}$$

где $\mathbf{N}^T$ — усилие, вызванное тепловым воздействием, отнесенное к единице длины линий, ограничивающих элемент рассматриваемой поверхности, Н/м; $\mathbf{M}^T$ — момент силы, вызванный тепловым воздействием, отнесенный к единице длины линий, ограничивающих элемент рассматриваемой поверхности, Н.

В рассматриваемом нами случае внешнее механическое нагружение отсутствует, поэтому уравнения (2, 3) можно переписать следующим образом:

$$\begin{pmatrix} N_x^T \\ N_y^T \\ N_{xy}^T \end{pmatrix} = \begin{pmatrix} A_{11} & A_{12} & A_{16} \\ A_{12} & A_{22} & A_{26} \\ A_{16} & A_{26} & A_{66} \end{pmatrix} \begin{pmatrix} \varepsilon_x^0 \\ \varepsilon_y^0 \\ \gamma_{xy}^0 \end{pmatrix} + \begin{pmatrix} B_{11} & B_{12} & B_{16} \\ B_{12} & B_{22} & B_{26} \\ B_{16} & B_{26} & B_{66} \end{pmatrix} \begin{pmatrix} \kappa_x \\ \kappa_y \\ \kappa_{xy} \end{pmatrix},$$

$$\begin{pmatrix} M_x^T \\ M_y^T \\ M_{xy}^T \end{pmatrix} = \begin{pmatrix} B_{11} & B_{12} & B_{16} \\ B_{12} & B_{22} & B_{26} \\ B_{16} & B_{26} & B_{66} \end{pmatrix} \begin{pmatrix} \varepsilon_x^0 \\ \varepsilon_y^0 \\ \gamma_{xy}^0 \end{pmatrix} + \begin{pmatrix} D_{11} & D_{12} & D_{16} \\ D_{12} & D_{22} & D_{26} \\ D_{16} & D_{26} & D_{66} \end{pmatrix} \begin{pmatrix} \kappa_x \\ \kappa_y \\ \kappa_{xy} \end{pmatrix}.$$

Преобразуем эту систему уравнений:

$$\varepsilon^0 = (\mathbf{A}^{-1} + (\mathbf{A}^{-1}\mathbf{B})(\mathbf{D} - \mathbf{B}\mathbf{A}^{-1}\mathbf{B})^{-1}(\mathbf{B}\mathbf{A}^{-1}))\mathbf{N}^T - (\mathbf{A}^{-1}\mathbf{B})(\mathbf{D} - \mathbf{B}\mathbf{A}^{-1}\mathbf{B})^{-1}\mathbf{M}^T, \tag{6}$$

$$\kappa = -(\mathbf{D} - \mathbf{B}\mathbf{A}^{-1}\mathbf{B})^{-1}(\mathbf{B}\mathbf{A}^{-1})\mathbf{N}^T + (\mathbf{D} - \mathbf{B}\mathbf{A}^{-1}\mathbf{B})^{-1}\mathbf{M}^T, \tag{7}$$

где $\mathbf{A}$ — матрица жесткости при растяжении (мембранная жесткость), Н/м; $\mathbf{B}$ — матрица жесткости изгиб-растяжение (смешанная жесткость), Н; $\mathbf{D}$ — матрица жесткости при изгибе (изгибная жесткость), Н·м.

Зависимости для определения коэффициентных напряжений в любой плоскости внутри рассматриваемой сборки параллельной срединной поверхности получаются подстановкой уравнений (4 — 7) в уравнение (1). Причем, зная расположение главных осей кремния, преобразуя значения элементов матриц жесткости в элементы матрицы $\mathbf{Q}$ в соответствии с формулами поворота системы координат [10], можно определять напряжения в любых направлениях.

**Примеры применения**

Исходя из данных о КТЛР применяемых стекла и кремния, на основании вышеописанной последовательности расчетов, можно определить значения коэффициентных напряжений при рабочей температуре $T_\text{р}$ в деталях, соединенных при температуре $T_\text{с}$. Проведя предварительный расчет, можно спланировать процесс соединения, обеспечивающий минимальные коэффициентные напряжения или же выдерживающий их в определенных пределах. Также, зная температуру, при которой провели соединение деталей, можно оценить характер изменения коэффициентных напряжений в рабочем диапазоне температур получаемого изделия. В связи со сложностью описанной методики расчета, рекомендуется проводить его при помощи ЭВМ.

Экспериментально полученная зависимость КТЛР кремния приведена в литературе [12]. Расчет зависимости КТЛР для стекол был проведен на основании данных, указанных производителями, исходя из предположения, что температурный ход значения коэффициента в области «нормального» состояния стекол, от минус 120 °C до температуры нижней границы зоны отжига, практически выражается линейным уравнением [4]. Для стекла ЛК5 [9]: модуль Юнга — 68,45 ГПа, КТЛР в диапазоне от минус 60 до плюс 20 °C составляет $33 \cdot 10^{-7}$ 1/°C, КТЛР в диапазоне 20…120 °C — $35 \cdot 10^{-7}$ 1/°C.

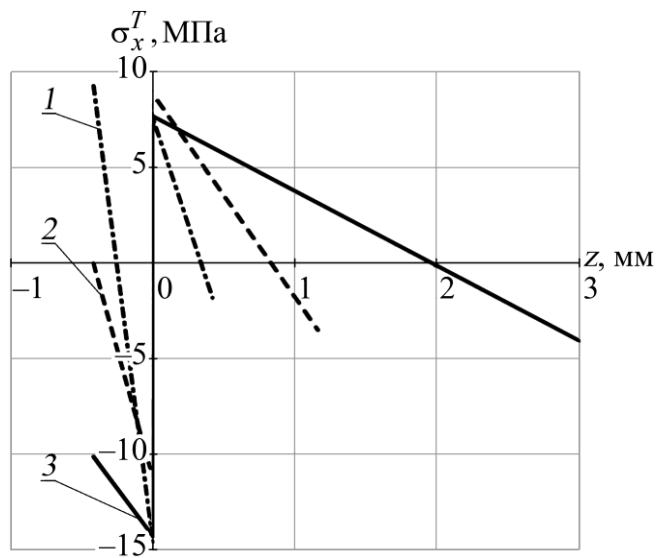

Рис. 2. Графики распределения коэффициентных напряжений в сборке при рабочей температуре 20 °C (температура соединения 440 °C), рассчитанные по двухслойной модели кремний-стекло, для случаев разных толщин стеклянного слоя.

На рисунке 2 представлены графики распределения коэффициентных напряжений в сборке при рабочей температуре 20 °C (температура соединения 440 °C), рассчитанные по двухслойной модели кремний-стекло, для случаев разных толщин стеклянного слоя. Марка стекла, использованная в расчете — ЛК-5. За плоскость отсчета координаты по оси $z$ взята плоскость соединения кремния со стеклом. Обозначения графиков: *1* — толщина пластин кремния и стекла равны 0,42 мм; *2* — толщина пластины кремния равна 0,42 мм, толщина пластины стекла равна 1,29 мм; *3* — толщина пластины кремния равна 0,42 мм, толщина пластины стекла равна 3 мм. Показательно, что напряжения, как в кремнии, так и в стекле могут менять свой знак на протяжении толщины материала. Также из данной иллюстрации можно сделать вывод что, варьируя толщину стекла, можно получить нулевые напряжения на некоторой глубине кремния или же иметь на этой глубине напряжения предсказуемого значения.

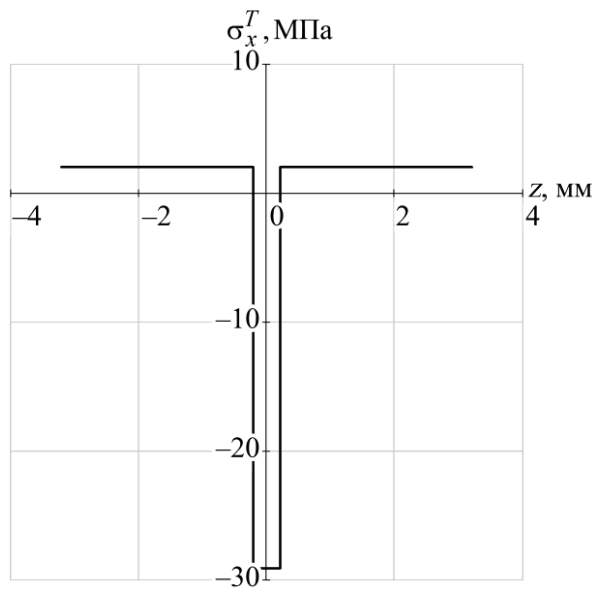

Рис. 3. График распределения коэффициентных напряжений в сборке при рабочей температуре 20 °C (температура соединения 450 °C), рассчитанные по трехслойной модели стекло-кремний-стекло.

На рисунке 3 представлен график распределения коэффициентных напряжений в сборке при рабочей температуре 20 °C (температура соединения 450 °C), рассчитанный по трехслойной модели стекло-кремний-стекло. Марка стекла, использованная в расчете — ЛК-5. За плоскость отсчета координаты по оси *z* взята срединная плоскость пластины кремния. Толщина пластины кремния равна 0,42 мм, толщина пластин стекла равна 3 мм. Следует отметить, что в таких симметричных структурах, как, например, стекло-кремний-стекло, согласно рассматриваемой модели, коэффициентные напряжения постоянны по толщине слоев. Причиной тому являются допущения применяемой модели расчета.

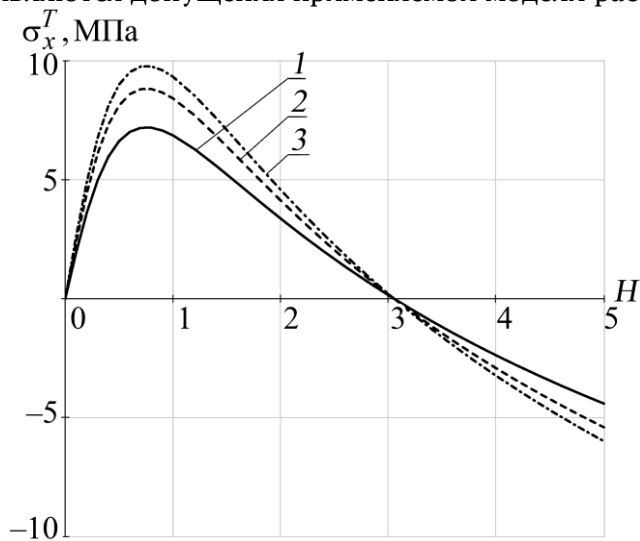

Рис. 4. Напряжения на свободной поверхности кремния при рабочей температуре 20 °C в зависимости от соотношения толщин стекла и кремния, рассчитанные по двухслойной модели кремний-стекло: *1* — температура соединения 300 °C; *2* — температура соединения 400 °C; *3* — температура соединения 450 °C.

На рисунке 4 представлены графики напряжений на свободной поверхности кремния при рабочей температуре 20 °C в зависимости от отношения, *H*, толщины стекла к толщине кремния рассчитанные по двухслойной модели кремний-стекло. Марка стекла использованная в расчете — ЛК-5. За плоскость отсчета координаты по оси *z* взята срединная плоскость пластины кремния. Толщина пластины кремния, взятая в расчетах, составляет 0,42 мм. Можно сделать вывод, что существует такое соотношение толщины стекла и кремния, при котором на поверхности кремния будут отсутствовать коэффициентные напряжения, независимо от температуры, при которой было проведено соединение. Расчетным путем установлено, что это соотношение составляет 3,05 для соединений стекла ЛК-5 и кремния.

**Экспериментальные результаты**

Были проведены эксперименты по соединению пластин кремния и стекла. Использовались пластины кремния КЭС диаметром 60 мм ориентации {100} с удельным сопротивлением 0,01 Ом·см и прямоугольные пластины стекла ЛК-5 размерами 30 × 50 × 4,5 мм. Соединения были проведены при температуре 330…350 °C. Затем были проведены измерения напряжений в стекле методом поляризационно-оптического измерения разности хода лучей, возникающей при прохождении через напряженное стекло линейно-поляризованного света и пропорциональной действующим напряжениям в стекле. Метод основан на явлении двулучепреломления [13], которое наблюдается в напряженном стекле при прохождении через него луча линейно-поляризованного света, и заключается в разложении луча на два — обыкновенный и необыкновенный, каждый из которых распространяется со своей скоростью. Вследствие этого эти лучи имеют при выходе из напряженного стекла разность хода. Измерения величины двулучепреломления в стекле проводились в соответствии с ГОСТ 3519-91 на полярископе-поляриметре ПКС-250.

В соединениях стекло-кремний наблюдалась смена знака напряжений в стекле от растягивающих к сжимающим по мере удаления от границы соединения, что хорошо соотносится с расчетными результатами, приведенными на рис. 2. После соединения стекло-кремний-стекло напряжения в стекле по мере удаления от плоскости соединения не изменялись, что можно считать качественной проверкой результатов моделирования, приведенных на рис. 3.

**Заключение**

В рамках данной работы рассмотрен способ расчета напряжений в сборках пластин стекла и кремния, соединенных методом электростатического соединения, проводимый в соответствии с теорией слоистых композитов. Проведены расчеты для некоторых случаев соотношений толщин и температур проведения процесса. Показано, что для симметричных структур напряжения по толщине соединяемых пластин не изменяются. Также сделан вывод о существовании такого соотношения толщин пластин стекла и кремния, при котором на поверхности кремния будут отсутствовать коэффициентные напряжения, независимо от температуры, при которой было проведено соединение.

**Список использованной литературы**
1. George Wallis, Daniel I. Pomerantz. Field Assisted Glass-Metal Sealing // Journal of Applied Physics. — 1969. — Vol. 40, No. 10. — P. 3946–3949.
2. J. Wei, H. Xie, M. L. Nai et al. Low temperature wafer anodic bonding // Journal of Micromechanics and Microengineering. — 2003. — Vol. 13, No. 2. — P. 217–222.
3. ОСТ 11 0041-85. Стекло электровакуумное. Термины и определения. — Введ. 1986-01-01. — 1985. — 42 с.
4. О. В. Мазурин, А. С. Тотеш, М. В. Стрельцина, Т. П. Швайко-Швайковская. Тепловое расширение стекла — Л.: Наука, Ленингр. отд-ние, 1969. — 216 с.: ил.
5. Р. Кристенсен. Введение в механику композитов / перевод с англ. А. И. Бейля и Н. П. Жмудя; под ред. Ю. М. Тарнопольского. — М.: «Мир», 1982. — 336 с.
6. A. Cozma, B. Puers. Characterization of the electrostatic bonding of silicon and Pyrex glass // Journal of Micromechanics and Microengineering. — 1995. — No. 5. — P. 98–102.
7. M. Bao. Analysis and Design Principles of MEMS Devices — Amsterdam: Elsevier, 2005. — 327 p.
8. В. И. Феодосьев. Сопротивление материалов: Учеб. для вузов — М.: Изд-во МГТУ им. Н. Э. Баумана, 1999. — 592 с.
9. LK5 glass type // Лыткаринский завод оптического стекла. Дата обновления: 07.07.2005. Систем. требования: Acrobat. URL: http://lzos.ru/glass_pdf/LK5.pdf (дата обращения: 09.10.2007).
10. Н. А. Алфутов Расчет многослойных пластин и оболочек из композиционных материалов / Н. А. Алфутов, П. А. Зиновьев, Б. Г. Попов — М.: Машиностроение, 1984. — 264 с., ил.
11. В. В. Васильев. Механика конструкций из композиционных материалов — М.: Машиностроение, 1988. — 272 с: ил.
12. H. Watanabe, N. Yamada, M. Okaji. Linear Thermal Expansion Coefficient of Silicon from 293 to 1000 K // International Journal of Thermophysics. — 2004. — Vol. 25, No. 1. — P. 221–236.
13. Стекло. Справочник / Под ред. Н. М. Павлушкина — М.: Стройиздат, 1973. — 487 с.: ил.

**References**
1. George Wallis, Daniel I. Pomerantz. Field Assisted Glass-Metal Sealing // Journal of Applied Physics. — 1969. — Vol. 40, No. 10. — P. 3946–3949.
2. J. Wei, H. Xie, M. L. Nai et al. Low temperature wafer anodic bonding // Journal of Micromechanics and Microengineering. — 2003. — Vol. 13, No. 2. — P. 217–222.
3. OST 11 0041-85. Steklo elektrovakuumnoe. Terminy i opredelenija. — Pub. 1986-01-01. — 1985. — 42 p.
4. O. V. Mazurin, A. S. Totesh, M. V. Strel'tsina, T. P. Shvajko-Shvajkovskaja. Teplovoe rasshirenie stekla — L.: Nauka, 1969. — 216 p.
5. R. Kristensen. Vvedenie v mehaniku kompozitov / translated by A. I. Bejlja and N. P. Zhmudja; editor Ju. M. Tarnopol'skij. — M.: «Mir», 1982. — 336 p.

6. A. Cozma, B. Puers. Characterization of the electrostatic bonding of silicon and Pyrex glass // Journal of Micromechanics and Micro-engineering. — 1995. — No. 5. — P. 98–102.
7. M. Bao. Analysis and Design Principles of MEMS Devices — Amsterdam: Elsevier, 2005. — 327 p.
8. V. I. Feodos'ev. Soprotivlenie materialov: Ucheb. dlja vuzov — M.: Izd-vo MGTU im. N. E. Baumana, 1999. — 592 p.
9. LK5 glass type // Lytkarinskij zavod opticheskogo stekla. Data obnovlenija: 07.07.2005. Sys. requirements: Acrobat. URL: http://lzos.ru/glass_pdf/LK5.pdf (data obrashhenija: 09.10.2007).
10. N. A. Alfutov Raschet mnogoslojnyh plastin i obolochek iz kompozitsionnyh materialov / N. A. Alfutov, P. A. Zinov'ev, B. G. Popov — M.: Mashinostroenie, 1984. — 264 p.
11. V. V. Vasil'ev. Mehanika konstruktsij iz kompozitsionnyh materialov — M.: Mashinostroenie, 1988. — 272 p.
12. H. Watanabe, N. Yamada, M. Okaji. Linear Thermal Expansion Coefficient of Silicon from 293 to 1000 K // International Journal of Thermophysics. — 2004. — Vol. 25, No. 1. — P. 221–236.
13. Steklo. Spravochnik / Editor N. M. Pavlushkina — M.: Strojizdat, 1973. — 487 p.